\documentclass[twocolumn,aps,prb,groupedaddress,showpacs]{revtex4}
\usepackage{graphicx}
\usepackage{amsmath,amssymb}
\usepackage{hyperref}
\usepackage{setspace}

\begin{document}

\title{Breakdown of self-averaging in the Bose glass}
\author{Anthony Hegg$^1$, Frank Kr\"uger$^2$, and Philip W. Phillips$^1$}  
\affiliation{$^1$Department of Physics, University of Illinois, 1110 West Green Street, Urbana, Illinois 61801, USA\\
$^2$SUPA, School of Physics and Astronomy, University of St. Andrews, St. Andrews, KY16 9SS, United Kingdom}
\date{\today}

\begin{abstract}
We study the square-lattice Bose-Hubbard model with bounded random on-site energies at zero temperature. Starting from a dual representation obtained from a strong-coupling expansion around the atomic limit, we employ a real-space block decimation scheme. This approach is non-perturbative in the disorder and enables us to study the renormalization-group flow of the induced random-mass distribution. In both insulating phases, the Mott insulator and the Bose glass, the average mass diverges, signaling short range superfluid correlations. The relative variance of the mass distribution distinguishes the two phases,  renormalizing to zero in the Mott insulator and diverging in the Bose glass. Negative mass values in the tail of the distribution indicate the presence of rare superfluid regions in the Bose glass. The breakdown of self-averaging is evidenced by the divergent relative variance and increasingly non-Gaussian distributions. We determine an explicit phase boundary between the Mott insulator and Bose glass.
\end{abstract}


\maketitle

\section{Introduction}

Introducing quenched disorder into an otherwise pure system can lead to subtle and complex results and the disordered 
Bose-Hubbard (BH) model is no exception. While the clean BH model shows a relatively straightforward bosonic competition between repulsion 
and tunneling, the disordered model exhibits a new gapless insulating phase, the Bose glass (BG), the precise location of which has proven
problematic from the outset.\cite{fisher1} With the advent of experimental methods that can engineer this model directly,\cite{greiner1}
the problem has been inverted and this has sparked renewed interest in the role disorder plays in quantum systems. 

\begin{figure}[t]
\centering
\includegraphics[scale=0.5]{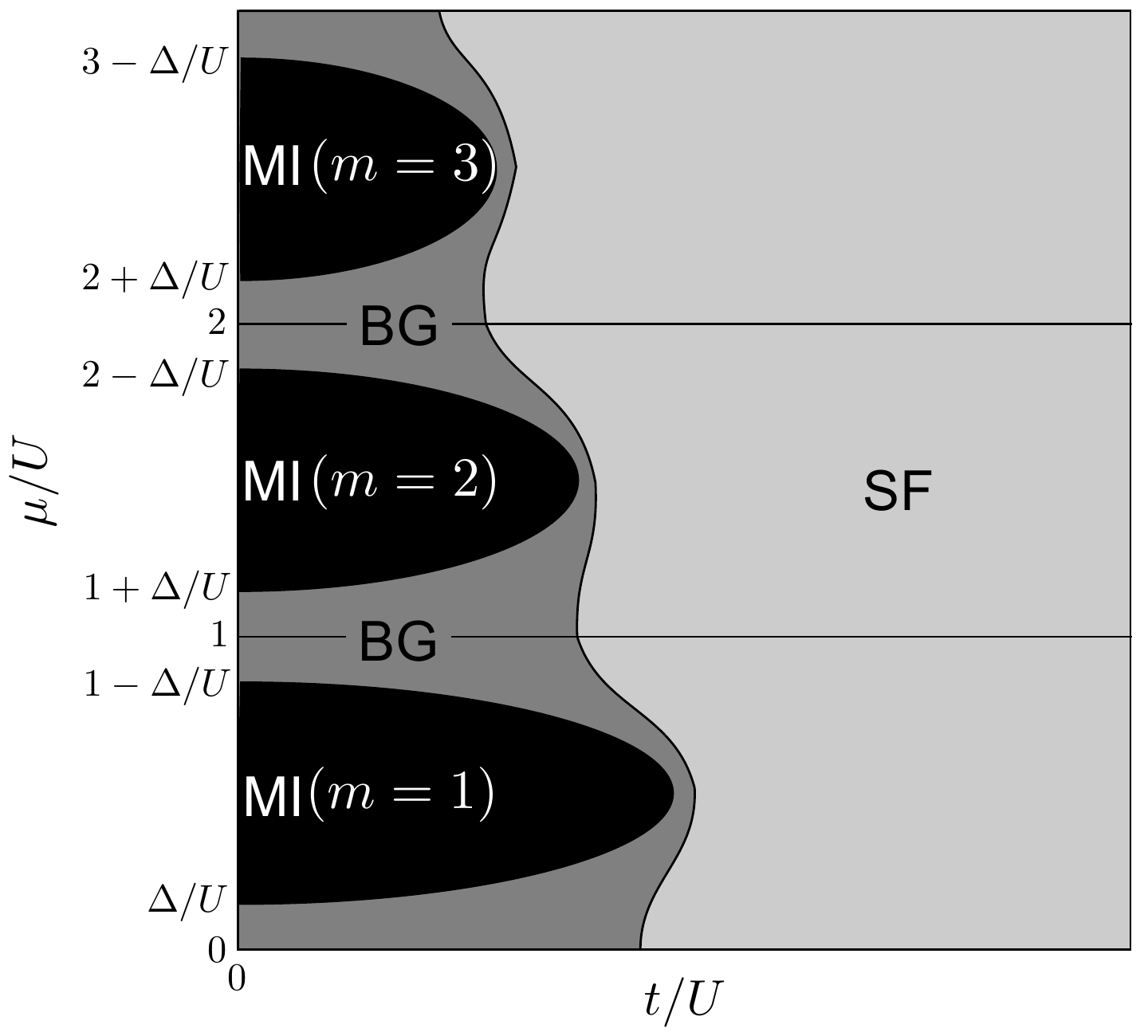}
\caption{Sketch of the phase diagram of the disordered Bose-Hubbard model as a function of the hopping $t/U$ and 
chemical potential $\mu/U$ in units of the on-site repulsion $U$. Random shifts in the on-site energies bounded by 
$\Delta<U/2$ lead to the formation of a compressible Bose-glass (BG) phase, separating the Mott-insulating (MI) lobes from the 
superfluid (SF).}
\label{fig:PD}
\end{figure}

In this Article, we develop a non-perturbative method to probe the nature of the transition between the localization-induced BG and 
the Mott insulator (MI) in the two-dimensional BH model with bounded potential disorder.  The MI arises from on-site 
repulsions and hence dominates in the limit the hopping vanishes while the superfluid (SF) is the ground state in the opposite
regime. It is in the difficult intermediate parameter space where the BG phase obtains.  It has been argued by various authors \cite{fisher1,kisker1,pollet1,gurarie,iyer1,niederle} that the BG is a quantum Griffiths phase dominated by arbitrarily large SF regions that are, 
however, exponentially suppressed. Despite the abundance 
of numerical \cite{scalettar1,krauth1,pai1,kisker1,sen1,lee1,prokofev1} and analytical \cite{singh1,wu1,bissbort1,mukhopadhyay1,freericks1,svistunov1,
herbut1,herbut2,weichman1,kruger1} work on the subject, it is only recently that several aspects of this model have been fully understood. This 
includes the confirmation that the BG always intervenes between MI and SF phases \cite{pollet1} (Fig.~\ref{fig:PD}), the proof that the transition 
between the MI and the BG has to be of the Griffiths type,\cite{gurarie} and the distinction between 
the MI and BG regarding whether fluctuations are self-averaging. \cite{kruger2}

As argued by Aharony and Harris,\cite{aharony1} the breakdown of self-averaging can be identified from the 
renormalization-group (RG) flow of the relative variance of any extensive variable. If the relative variance does not renormalize to 
zero, the central limit theorem no longer applies and the system is not self-averaging.  This concept has been used to characterize 
the phase transition between the MI and the BG \cite{kruger2} within a disorder averaged replica field theory. In both insulating phases, 
the mass of the theory diverges, signaling the presence of short-ranged SF 
correlations. In dimensions $d<4$ the variance of the mass distribution diverges as well and as a consequence, the breakdown of self-averaging can be readily 
understood as a competition between the spread of the distribution versus the shift of its average.  In the MI, the shift dominates the spread 
leading to a vanishing relative variance. In the BG, the spread is faster and the relative variance diverges. This characterization of the 
MI/BG transition is consistent with the picture that the BG is dominated by rare SF regions since the negative mass values
occur in the tail of the distribution. Whether the onset of this Griffiths instability \cite{pollet1,gurarie} is correctly described  
by a perturbative RG calculation remains a central question.
 
We address this question within a real-space block decimation RG scheme on a square lattice. This approach is 
non-perturbative in the disorder and hence enables a study of the RG flow of the full random-mass distribution, a necessity for any definitive 
statement about Griffiths-type physics to be made. We reiterate that
the onset of the Bose glass is well known to be mediated\cite{pollet1,gurarie} by
Griffiths rare-region physics and hence our conclusions are independent of whether the
transition is studied from the Mott insulator or the superfluid. Our results confirm that the relative variance serves as the order parameter for the 
MI-to-BG transition. Determining the correlation length from the scale at which the relative variance becomes of order one, we extract a 
correlation-length exponent of about $\nu=0.7$. This is close to the analytical value $\nu=1/d$ obtained within the perturbative 1-loop 
RG.\cite{kruger2} It has been argued \cite{aharony1,pazmandi1} that a violation of the Harris-criterion bound $\nu \geq 2/d$ for critical 
disordered systems \cite{harris1,chayes1} is indicative of the lack of self-averaging. The absence of the central-limit 
theorem in the BG is further evidenced by an increasingly non-Gaussian shape of disorder distributions.

\section{Model}

Our starting point is the simplest form of the disordered BH model on a square lattice, 
\begin{equation} \label{eqn:BHH}
\mathcal{H}  =   -t \sum_{\langle i,j \rangle} \left( b^{\dag}_{i} b_{j} + b^{\dag}_{j} b_{i} \right) - \sum_{i} \mu_{i} \hat{n}_{i}
+\frac{U}{2} \sum_{i} \hat{n}_{i} \left(\hat{n}_{i} - 1 \right),
\end{equation}
which describes bosons tunneling with amplitude $t$ between nearest-neighbor sites $i$ and $j$ and interacting via an on-site 
repulsion $U$. The bosonic raising and lowering operators are given by $b_i^{\dag},b_i$ respectively where $\hat{n}_{i}=b^{\dag}_{i} b_{i}$ 
is the bosonic number operator. $\mu_i$ is the chemical potential shifted by the on-site disorder potential, $\mu_i=\mu-\epsilon_i$.
The random site energies $\epsilon_i$  are uncorrelated between different sites and uniformly distributed in the interval $[-\Delta,\Delta]$. 
From minimization of the energy in the atomic limit it is straightforward to see that for $\Delta<U/2$ the phase diagram retains MI phases
(see Fig.~\ref{fig:PD}).

To facilitate a strong coupling expansion, we follow the standard procedure.\cite{sachdev1,sengupta1}
After expressing the model by a coherent-state path integral in imaginary time, we decouple the hopping term by a Hubbard-Stratonovich transformation
and trace over the original boson fields. We then perform a temporal gradient expansion to obtain the effective dual action on a lattice,
\begin{eqnarray} \label{eqn:EffAct}
S_{\text{eff}} & = &  a^2  \sum_{i} \int d\tau \left( \frac{1}{2}\sum_{\delta}T_{i\delta}|\psi_{\delta}
-\psi_{i}|^2+K^{(1)}_i \psi^{*}_{i}\partial_{\tau} \psi_{i}\right. \nonumber\\
& & \left.  \vphantom{\frac{|\psi_{\delta}-\psi_{i}|^2}{a^2}} + K^{(2)}_i |\partial_{\tau} \psi_{i}|^2 + R_{i}|\psi_{i}|^2 + H_{i}|\psi_{i}|^4 \right),
\end{eqnarray}
where $a$ denotes the lattice spacing and the sum over $\delta$ runs over the nearest neighbors of site $i$. By construction, the complex fields 
correspond to the SF order parameter $\psi_{i}\sim\langle b_i\rangle$. In regions where $R_i>0$, SF order is suppressed. The mass $R_i$ 
therefore corresponds to the local Mott gap. $R_i$ and $H_i$ are related
to the single- and two-particle bosonic Green functions of the local on-site Hamiltonian, respectively, while the temporal gradient terms $K^{(1)}_i$ and 
$K^{(2)}_i$ are given by derivatives of the mass with respect to the
chemical potential. We specialize to the first Mott lobe with $m=1$
bosons per site in which
the coefficients are given by \cite{sengupta1,kruger1}
\begin{subequations}\label{eqn:initial}
\begin{eqnarray} \label{eqn:initial_mass}
R_i & = &  \frac{1}{zt} - \left( \frac{2}{U -\mu_{i}} + \frac{1}{\mu_{i}} \right),\\
K^{(1)}_i & = & -\frac{\partial R_i}{\partial\mu_i}, \quad K^{(2)}_i=-\frac 12 \frac{\partial^2 R_i}{\partial\mu_i^2},\\
H_i & = & \left( \frac{2}{U -\mu_{i}} + \frac{1}{\mu_{i}}  \right)\left( \frac{2}{(U -\mu_{i})^2} + \frac{1}{\mu_{i}^2} \right)\nonumber\\
& & -\frac{6}{(U -\mu_{i})^2(3U-2\mu_i)},
\end{eqnarray}
\end{subequations}
where $z=2d=4$ is the coordination number of the square lattice. In the clean limit, $\mu_i=\mu$, the mean-field phase boundary between the first 
MI lobe and the SF is obtained from $R(\mu,t,U)=0$. In the presence of disorder, $\mu_i=\mu-\epsilon_i$, the coefficients $R_i$, 
$K^{(1)}_i$, $K^{(2)}_i$, and $H_i$ depend on the disorder potential $\epsilon_i$, which induces non-trivial disorder distributions of the coefficients.
Note that initially the dual hopping amplitudes $T_{i\delta}=1/(za^2 t)$ are uniform. We allow for a spatial dependence of the hopping since disorder 
will be  induced under block decimation.

\section{Block Decimation Real-Space RG}

\begin{figure}[h]
\centering
\includegraphics[width=0.55\linewidth]{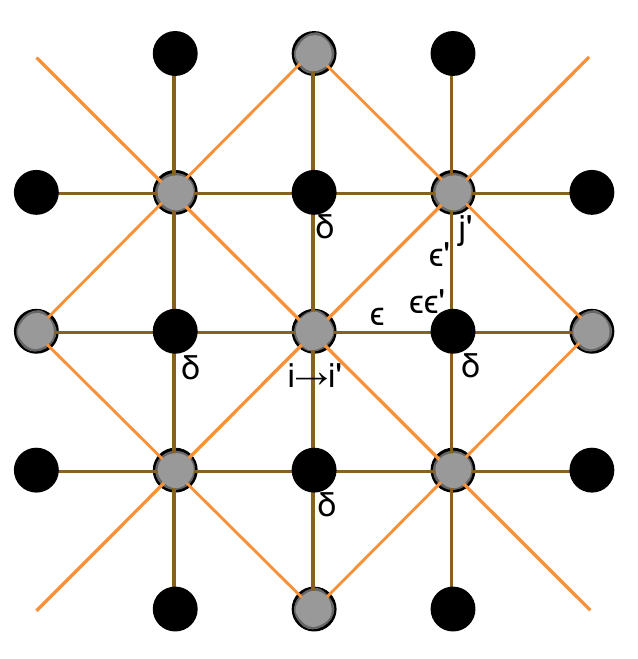}
\caption{(Color online) Illustration of the block decimation scheme underlying the recursion relations (\ref{eqn:RG}).}
\label{fig:BlockDecimation}
\end{figure}

Equations (\ref{eqn:initial}) provide the initial conditions for our procedure where the random on-site energies are generated from a uniform 
distribution on the interval $[-\Delta,\Delta]$.  In order to determine the phase for a given set of parameters $t$, $U$, $\mu$, and $\Delta$ 
of the disordered BH model (\ref{eqn:BHH}), we derive a set of recursion equations using block decimation. We start with the discrete action 
(\ref{eqn:EffAct}) and eliminate short-range degrees of freedom by integrating out every other 
site, treating the quartic terms $H_i$ perturbatively to leading order.  The remaining points form a new square lattice with lattice spacing 
$a'=\sqrt{2}a$ and tilted $45^{\circ}$ from the original system.    Recollecting the resulting 
terms and rescaling the action to look like the original, we find the RG recursion equations
\begin{subequations}\label{eqn:RG}
\begin{eqnarray}
R'_{i'} &=& R_{i} + \sum_{\delta}T_{i\delta} - \sum_{\delta,\delta'}T_{i\delta}T_{i\delta'}I_\delta\left( 1 - 4H_{i}\tilde{I}_\delta I_\delta \right),\quad\\
T'_{i'j'} &=& \sum_{\epsilon,\epsilon'}T_{\epsilon}T_{\epsilon'}I_{\epsilon\epsilon'}\left( 1 - 4H_{\epsilon\epsilon'}\tilde{I}_{\epsilon\epsilon'}
I_{\epsilon\epsilon'} \right),\label{eqn:RG_T}
\end{eqnarray}
\end{subequations}
where $I_i=( R_{i}+\sum_{\delta}T_{i\delta})^{-1}$ is the static propagator and $\tilde{I}_i=\left( 4( R_{i}+\sum_{\delta}T_{i\delta} )
K^{(2)}_i+(K^{(1)}_i)^2 \right)^{-\frac{1}{2}}$. The indices $\delta$ and 
$\delta'$ correspond to nearest neighbors of site $i$, whereas $\epsilon$ and $\epsilon'$ correspond to the 
bonds adjacent to the bond connecting the sites $i'$ and $j'$ of the remaining lattice. The site $\epsilon\epsilon'$ is the common vertex of the
bonds $\epsilon$ and $\epsilon'$ (see Fig.~\ref{fig:BlockDecimation}). Since we are interested in the MI/BG transition at incommensurate filling, 
we can neglect any corrections to the coefficients $K^{(1)}$, $K^{(2)}$, and $H$ beyond dimensional scaling. Note that under the RG longer 
range couplings are generated which is a known problem of the block decimation method in $d>1$. In the present case, however, locality is guaranteed since 
the mean of the mass distribution diverges in both insulating phase, leading to an exponential suppression of hopping amplitudes beyond nearest 
neighbors.

Since the gradient terms are renormalized under block decimation the effective local mass should 
be defined relative to the kinetic hopping terms, 
\begin{equation}
r_{i}:=z R_{i}/\sum_{\delta}T_{i\delta}.
\end{equation}

\begin{figure}[t]
\centering
\includegraphics[width=\linewidth]{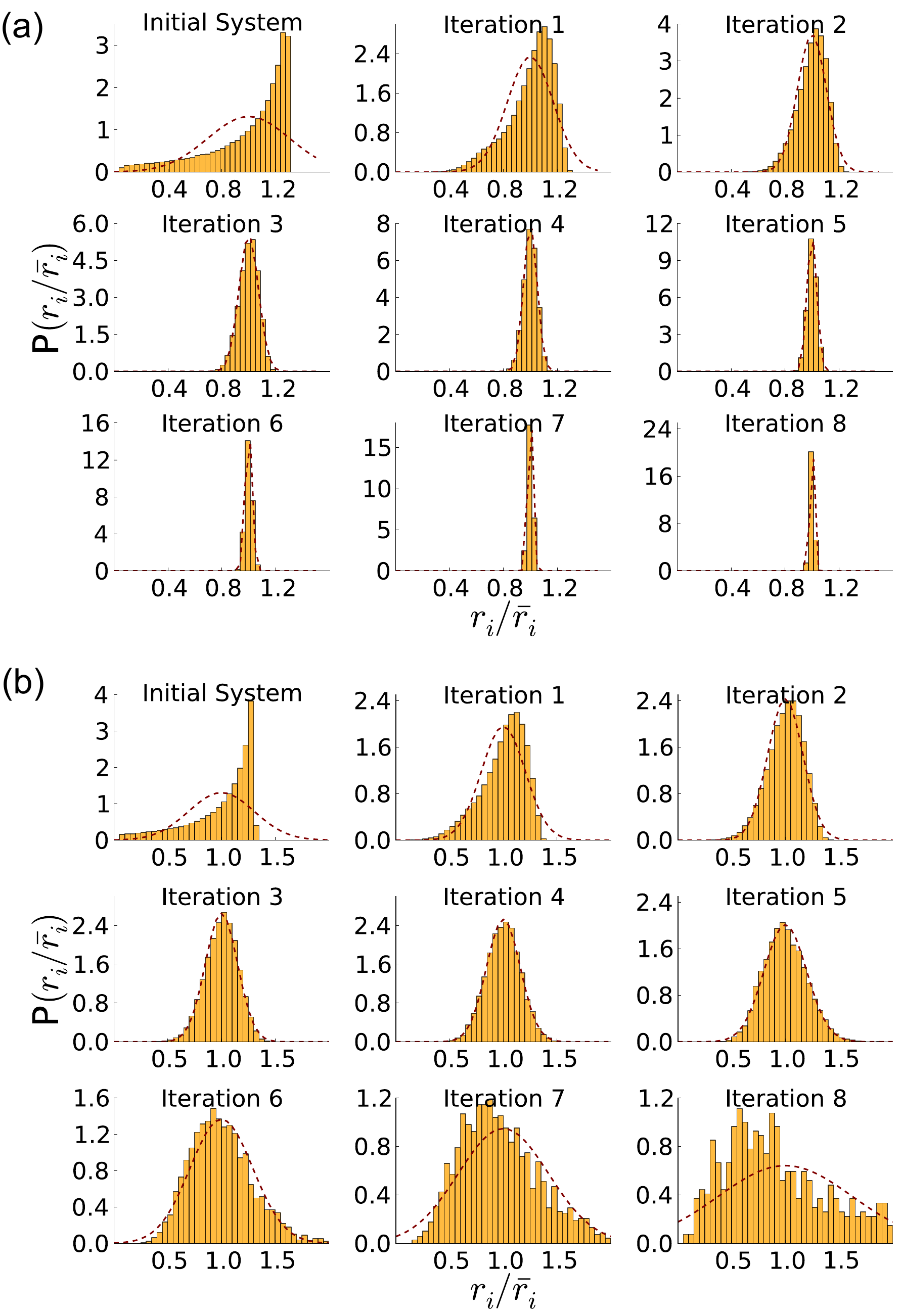}
\caption{(Color online) RG flow of the mass distributions $P(r_{i}/\overline{r_i})$ (normalized to a mean of $1$).
(a) In the MI, the Gaussian fit (red dotted line) becomes better with successive iterations and
the width of the distribution, which corresponds to the relative variance, narrows. (b) In the BG, the initial distribution is shifted to the left due to a larger 
initial hopping value. The Gaussian fit becomes worse and worse and the relative variance increases.}
\label{fig:Hist}
\end{figure}

As a consistency check, we evaluate our recursion relations in the clean limit. For $T=T_{i\delta}$, $R=R_i$, and $H=H_i=0$, the RG equations 
(\ref{eqn:RG}) reduce to $R'=R+4T-16T^2/(R+4T)$ and $T'=4T^2/(R+4T)$, leading to the recursion relation $r'=2r+r^2/4$ for the 
effective mass. This indeed correctly describes the mean-field transition between the MI and the SF at $r=0$.

\section{Results}

In the following, we integrate the RG equations (\ref{eqn:RG}) numerically for the inhomogeneous system obtained for one particular disorder 
realization and keep track of the values of the coupling constants on each lattice site. This allows us to extract the mass distribution $P(r_i)$ 
at each iteration step of the block decimation. We vary $t/U$ while keeping the disorder fixed at $\Delta=0.1$ for chemical potential values $\mu/U=0.15,0.2,...,0.75,0.8$ and several lattice constants $a$.
Since Eq.~(\ref{eqn:EffAct}) has only nearest neighbor or on-site terms, we can ignore the boundary 
points after each iteration without affecting the overall distribution. The major limitation to this method is the necessity of finite size lattices. 
Estimates of any diverging quantities near the critical point must take into account finite system size effects.  In the data given below we use an 
initial square grid of points with side length $L=506$ sites. This side length is not a power of two because each decimation step concludes by throwing away points affected by the boundary.

As expected for insulating phases, in both the MI and the BG, the mean $\overline{r_i}$ of the mass distribution increases exponentially under the RG
signaling short-range SF correlations. To distinguish the behavior in the MI and the BG, we normalize to a mean of unity and analyze the evolution of 
$P(r_i/ \overline{r_i})$. The variance of this rescaled distribution corresponds to the relative variance of the mass distribution,
which should serve as order parameter.\cite{kruger2}

In Fig.~\ref{fig:Hist}a, the RG flow of $P(r_i/ \overline{r_i})$ in the MI phase is shown. Note that the initial mass distribution is asymmetric
and non-Gaussian due to the functional dependence (\ref{eqn:initial_mass}) on the uniformly distributed on-site energies $\epsilon_i$. 
As a consequence of the relatively small hopping value $t/U$,  the bulk of the sites begin well above $r_{i}=0$ and 
continue together towards larger values under repeated iteration. 
Increasing the number of iterations results in a distribution well
described by a Gaussian.  Further, the width of the distribution narrows, indicating a vanishing relative variance. This demonstrates that in the MI 
disorder is irrelevant and the system is self-averaging.

\begin{figure}[t]
\centering
\includegraphics[width=\linewidth]{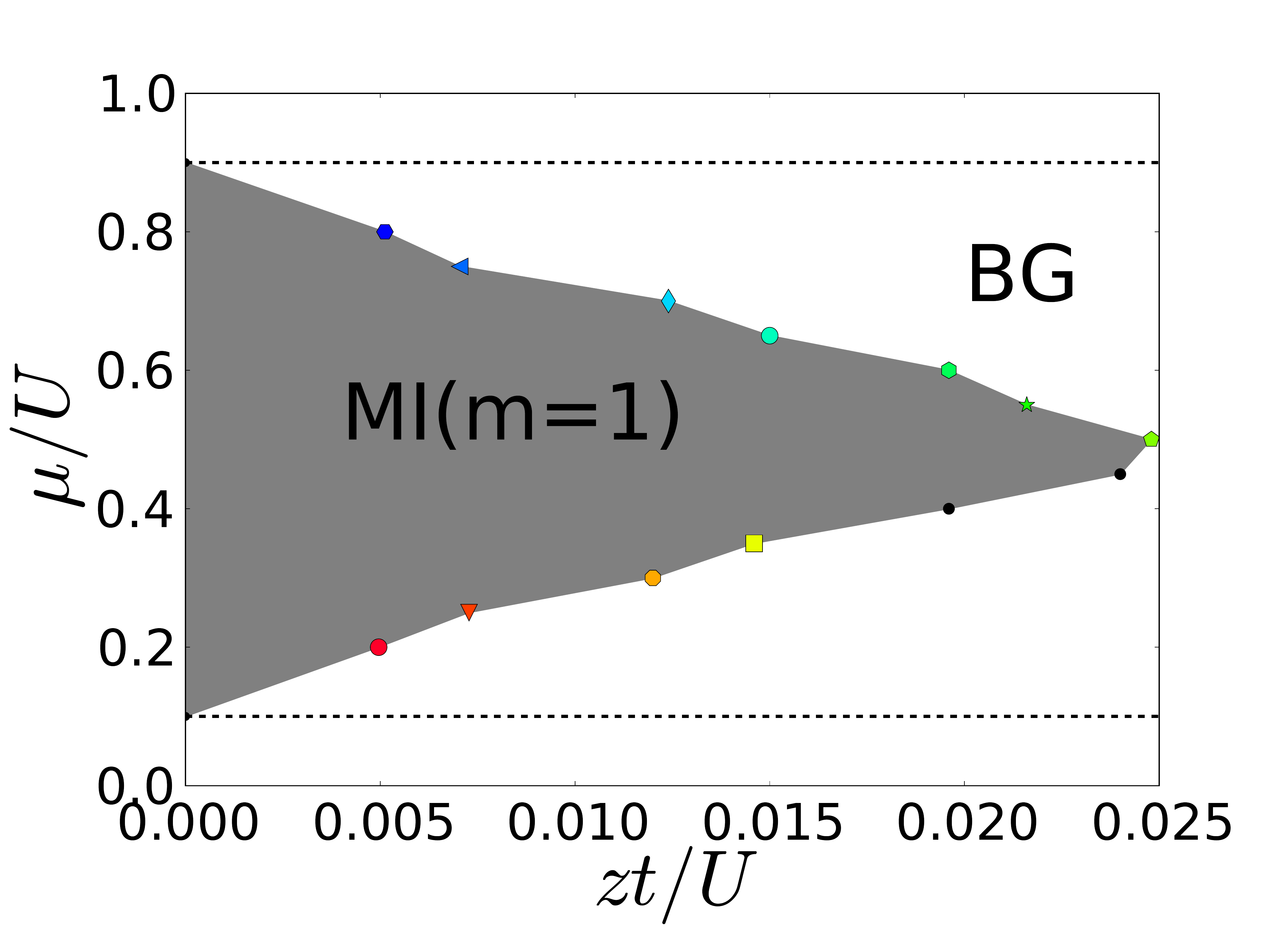}
\caption{(Color online) MI/BG phase boundary for the first Mott
  lobe obtained using the relative variance of the disorder-induced
  mass distribution as an order parameter. The critical hopping values $t_{c}$ and the shape of the phase boundary are
  on the order of those obtained via various computational methods.\cite{niederle} 
  The dotted lines correspond to $\mu/U=0.1,0.9$. The
  Mott insulator occurs only between the dotted lines.  The distance
  between the dotted lines and the integer-value fillings is
  set by the disorder width, $\Delta$. Note that the absolute values
  for $t_{c}$ vary with cutoff $a$, so only the relative shape of the diagram and an order of magnitude estimate of these values are obtained here.}
\label{fig:CritPD}
\end{figure}

The situation changes dramatically as we increase the value of $t/U$ and enter the BG phase (see Fig.~\ref{fig:Hist}).
While the overall shape of the initial distribution looks quite similar to the one in the MI, now a large fraction of the initial sites lies close to or 
below $r_{i}=0$. With enough sites close to the transition point, some regions of the system take much longer to flow to large positive values than 
others. This results in a drastic spread of values upon repeated iteration of the RG, leading to a divergence of the relative variance. 
In addition, the distribution develops a non-Gaussian form under successive iterations, indicating a violation of the central limit theorem. 
This demonstrates a breakdown of self-averaging in the BG. 

\begin{figure}[t]
\centering
\includegraphics[width=\linewidth]{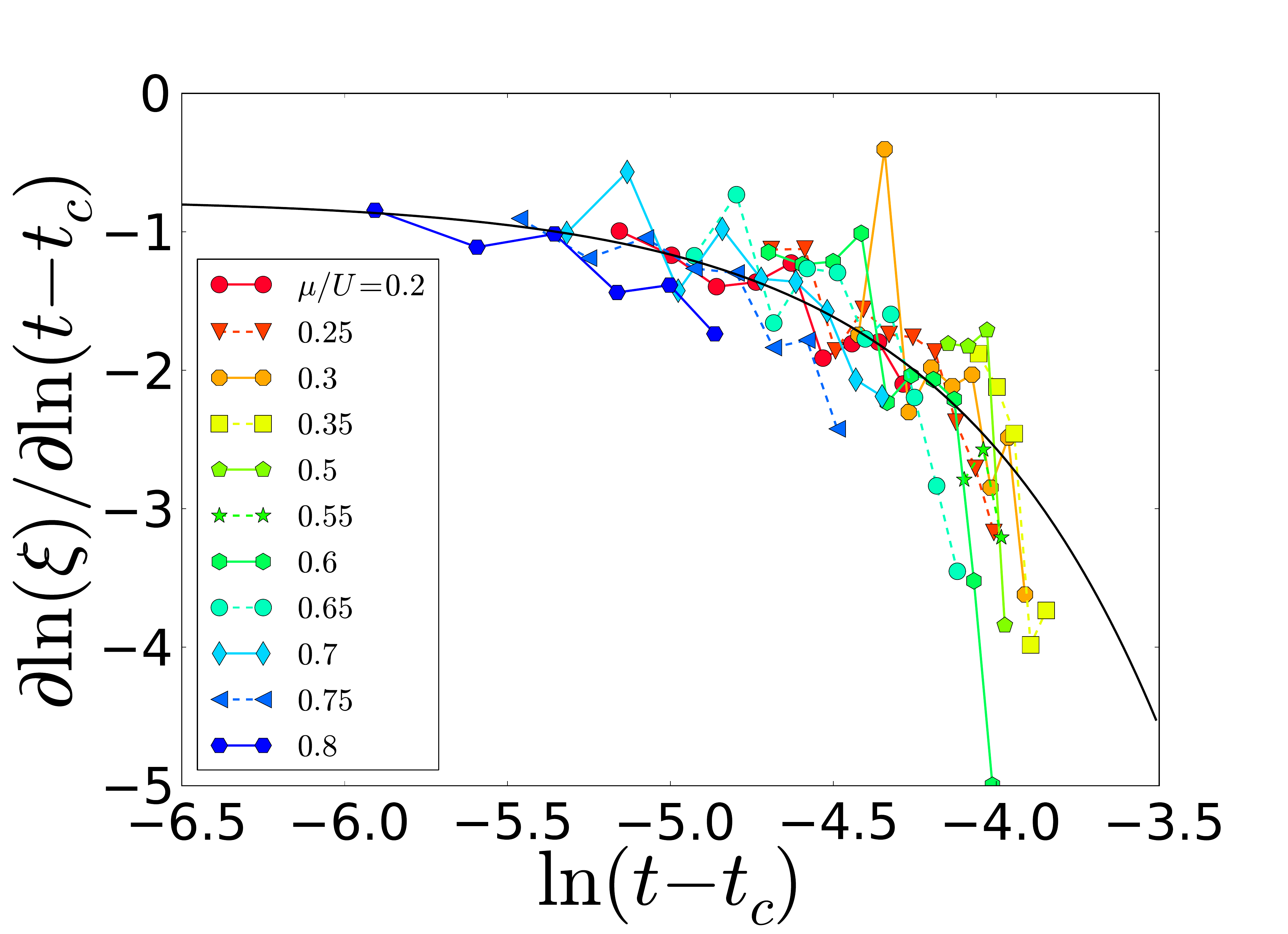}
\caption{(Color online) Scale collapse of the correlation length
  curves as a function of the hopping $t$ for various values of
  $\mu/U$. The curves' colors and markers are matched to
  Fig.~\ref{fig:CritPD}. Universality is seen at smaller scales
  further from the tip of the Mott lobe as is shown in previous
  momentum-shell methods.\cite{kruger2} This is indicated here by the
  higher and lower values of $\mu/U$ tending towards a universal
  constant value. For fillings slightly less than commensurate, the
  system size was too small to produce an estimate of $\nu$. The
  nonlinear fit predicts an asymptotic value of $\nu\approx 0.7$ for the correlation length exponent.}
\label{fig:CorrFit}
\end{figure}

Our results show that the relative variance of the random-mass distribution $r_i$ serves as an order parameter for the transition between the 
MI and the BG. We can therefore determine a correlation length $\xi/a=\sqrt{2}^{n_c}$ in the BG from the average number of iterations it takes
before the relative variance becomes of order unity. For the parameters used in Fig.~\ref{fig:Hist}b this happens between 6 and 7 iterations.
A more precise value is obtained by averaging over several disorder realizations. Note that once the relative variance becomes of order unity, the 
left tail of the distribution pushes through zero. Therefore, the correlation length $\xi$ corresponds to the 
typical distance between SF droplets in the system. 

We generate a phase diagram for the transition between the MI and BG
by estimating values for $t_{c}$ such that the data for several
initial conditions collapse onto a single curve. This method, and
consequently the resulting phase diagram, can only predict the
relative value of $t_{c}$ between different values of $\mu/U$ and not
the absolute location of $t_{c}$. Therefore, we set $t_{c}$ to be a
small constant value for initial conditions in which the BG is
suppressed and plot the results for intermediate values in
Fig.~\ref{fig:CritPD}. Other omputational methods have produced a
phase boundary on the same order of magnitude with the same
characteristic shape \cite{niederle}. 

At the transition to the MI, the
correlation length diverges as a power law, $\xi \sim(t-t_{c})^{-\nu}$.  To extract the correlation-length 
exponent $\nu$, we vary the hopping slightly above the transition point for fixed values of the lattice spacing $a$ and several values of the chemical potential $\mu/U$
and extract the correlation length as described above by averaging over several disorder realizations. In the following, we use $a=0.3$, $\mu/U=0.15, 0.2,...,0.75,0.8$ and average over 10-20 disorder realizations. 
Note that the value $t_c$ of the transition point is
non-universal. With the values of $t_{c}$ in Fig.~\ref{fig:CritPD}, we
plot  $\partial\ln\xi/\partial\ln(t-t_{c})$ vs $\ln(t-t_c)$ for
various values of $\mu/U$.  
The data collapse onto a single curve with an asymptotic value of $\nu
\approx 0.7$ as shown in Fig.~\ref{fig:CorrFit}. Values of the
chemical potential near commensurate density require larger
system sizes to see universal behavior as was found in
previous momentum-shell work.\cite{kruger2} This is indicated in
Fig.~\ref{fig:CorrFit} where the mid-range values of $\mu/U$ are
further from the universal asymptotic fit than those at the
extrema. As a result,  the system sizes used here of $L=506$ 
are too small to obtain enough data to estimate $\nu$ for filling values slightly 
less than commensurate.

\section{Discussion and Conclusion}

We used a real-space block decimation method to characterize the MI-to-BG transition. By analyzing the RG 
flow of the induced random-mass distribution in the dual field theory, we have demonstrated that 
the transition is characterized by a breakdown of self-averaging. The associated correlation length corresponds to the typical separation of rare SF 
regions. Note that while the transition is not related to spontaneous symmetry breaking, the divergence of the correlation length defined here has
been shown to be related to a vanishing compressibility.\cite{kruger2} Our work provides an explicit confirmation that the instability of the MI 
towards the formation of the BG is of the Griffith type, as previously argued by other authors.\cite{pollet1,gurarie} 

The method employed here enables us to study the RG flow of entire 
disorder distributions, whereas the perturbative 1-loop momentum-shell RG based on the disorder averaged replica theory is restricted to 
the mean and the variance of the random mass distribution.\cite{kruger2} Both approaches, however, show that the relative variance diverges in the 
BG. This is an important result as it demonstrates that contrary to the general belief, the onset of Griffiths instabilities is captured in perturbative RG. 
The comparison of the correlation-length exponents obtained by the two methods suggests that corrections beyond 1-loop order are small.

Griffiths phases can been classified based on the comparison of the dimension of defects (or rare regions) with the lower 
critical dimension.\cite{vojta2,vojta3,hoyos,mohan} Since the SF droplets are one-dimensional -- rod-like in imaginary time and zero dimensional in space --  the 
rare regions are \emph{below} the lower critical dimension of the problem, leading to weak Griffiths singularities characterized by an essential singularity in 
the free energy.\cite{classical} 

Our results show that the MI/BG transition is characterized by a fixed point with \emph{finite} relative variance. This is consistent with strong disorder RG 
calculations  that show that the transition between the SF and the disordered insulator is governed by a finite disorder 
fixed point.\cite{altman1,altman2,iyer1} While the strong disorder RG approach becomes asymptotically exact in the limit of infinite disorder, 
it might produce unphysical results in the regime of weak disorder.\cite{iyer1} Another difference to the block decimation scheme is that the lattice 
coordination is not preserved in dimensions $d>1$. It would be interesting to systematically compare the evolution of disorder distributions for the 
two complementary methods. 

The real-space RG approach presented here has a wide range of future applications. It can be used to study the effects of spatial correlations
in the disorder and entails the search for self-similar disorder characterized by scale invariant distribution functions. Finally, the 
method is not restricted to disorder distributions, but can be used to study other inhomogeneities such as the so-called wedding cake
structures of alternating MI and SF regions found in optical-lattice systems.\cite{Gemelke+09}

\textbf{Acknowledgment:}  The authors benefited from stimulating discussions with B. Brinkman, A.~V. Chubukov, I.~F. Herbut, J.~A. Hoyos, J. Keeling, R. Moessner, 
and H. Rieger. This work was supported by EPSRC under grant code EP/I 004831/1 and by NSF-DMR-1104909.


\begin{thebibliography}{38}
\bibitem[{cla()}]{classical}
\bibinfo{note}{Note that such weak Griffiths singularities are sometimes called
  `classical', see e.g. Ref.~[\onlinecite{vojta2}]. Although this case is
  indeed generic to classical systems with point disorder, the terminology
  `classical' is misleading since the BG is a $T=0$ phase.}
\end{thebibliography}
\end{document}